\documentclass{aa}
\usepackage{graphicx}

\newcommand{\be}{\begin{equation}}
\newcommand{\ee}{\end{equation}}

\def\ac{{\rm acc}}
\def\ad{{\rm adv}}
\def\ag{{\rm age}}
\def\l{{\rm loss}}
\def\di{{\rm dif}}
\def\co{{\rm cool}}
\def\u{{\rm u}}
\def\d{{\rm d}}
\def\e{{\rm e}}
\def\s{{\rm s}}

\def\f{\frac}

\def\Half{\frac{1}{2}}

\def\N{\nonumber}

\def\BE{\begin{equation}}
\def\EE{\end{equation}}
\def\BEA{\begin{eqnarray}}
\def\EEA{\end{eqnarray}}
\def\L{\label}
\def\MG{\mu{\rm G}}
\def\TeV{{\rm TeV}}
\def\EM{E_{\rm max}}
\def\max{{\rm max}}
\def\min{{\rm min}}
\def\NRO{\nu_{\rm rolloff}}

\begin{document}

\title{
Constraints On the Diffusive Shock Acceleration From 
the Nonthermal X-ray Thin Shells In SN~1006 NE Rim
}

\subtitle{}

\author{Ryo~Yamazaki\inst{1} 
\and Tatsuo~Yoshida\inst{2} 
\and Toshio~Terasawa\inst{3} 
\and Aya~Bamba\inst{1} 
\and Katsuji~Koyama\inst{1}
}

\offprints{R.~Yamazaki}

\institute{
Department of Physics, Kyoto University, Kyoto 606-8502, Japan\\
\email{yamazaki@tap.scphys.kyoto-u.ac.jp} ,
\email{bamba@cr.scphys.kyoto-u.ac.jp} ,\\
\email{koyama@cr.scphys.kyoto-u.ac.jp}
\and 
Faculty of Science, Ibaraki University, Mito 310-8512, Japan\\
\email{yoshidat@mx.ibaraki.ac.jp}
\and
Earth \& Planetary Science, Graduate School of Science, 
University of Tokyo, 7-3-1 Hongo, Bunkyo-ku, 113 Tokyo, Japan\\
\email{terasawa@eps.s.u-tokyo.ac.jp}
}
\date{Received date/ Accepted date}

\authorrunning{Yamazaki et al.}
\titlerunning{Magnetic field configuration in SN~1006 NE Rim}

\abstract{
Characteristic scale lengths of nonthermal X-rays from 
the SN~1006 NE rim, which are observed by {\it Chandra},
are interpreted in the context of the diffusive shock acceleration
on the assumption
that the observed spatial profile of nonthermal X-rays
corresponds to that of accelerated electrons with energies of a few
tens of TeV.
To explain the observed scale lengths,
we construct two simple models
with a test particle approximation,
where the maximum energy of accelerated electrons is determined
by the age of SN~1006 ({\it age-limited model})
or the energy loss ({\it energy loss-limited model}),
and constrain the magnetic field configuration
and the diffusion coefficients of accelerated electrons.
When the magnetic field is nearly parallel to the shock normal,
the magnetic field should be in the range of 20--85~$\MG$ and
highly turbulent both in upstream and downstream,
which means that the mean free path of accelerated electrons
is on the order of their gyro-radius (Bohm limit).
This situation can be realized both in the 
age-limited and energy loss-limited model.
On the other hand,
when the magnetic field is nearly perpendicular 
to the shock normal, which can exist only in the age-limited case,
the  magnetic field is several $\MG$ in the upstream and
14--20 $\MG$ in the downstream, and the upstream magnetic field 
is less turbulent than the downstream.



}\maketitle

\section{Introduction}

Galactic cosmic rays with energies of less than $10^{15.5}$~eV
(the ``knee'' energy) are commonly believed to be generated by
supernova remnants (SNRs).
SN~1006 is one of the SNRs which is thought to be an accelerator
of such high energy particles.
Koyama et al. (1995) discovered synchrotron X-rays from the rims 
of this SNR,
indicating the existence of 
accelerated electrons with an energy more than a few tens of TeV.
The detection of TeV $\gamma$-rays from the northeastern (NE) rim 
of SN~1006 (Tanimori et al. 1998) implies further evidence
for the presence of high energy particles, 
since TeV $\gamma$-rays arise from the Inverse Compton (IC) process,
in which cosmic microwave background (CMB) photons are 
up-scattered by high energy electrons (Tanimori et al. 2001), 
or the hadronic process,
in which $\pi^0$ particles made by collisions between
accelerated and interstellar protons decay into $\gamma$-ray photons
(Berezhko, Ksenofontov, \& V\"{o}lk 2002; Aharonian \& Atoyan 1999).

The mechanism for cosmic ray acceleration has also been studied
for a long time and the most plausible process is a diffusive 
shock acceleration (DSA)
(Bell 1978; Blandford \& Ostriker 1978; Drury 1983; 
Blandford \& Eicher 1987; Jones \& Ellison 1991;
Malkov \& Drury 2001).
Many authors have explained the observed properties of 
SN~1006 in the context of the DSA but
the conclusions are different due to the arbitrary assumptions
on unknown physical parameters, 
such as the magnetic field configuration,
the diffusion coefficient, the injection rate, and 
the electron to proton ratio
(Ellison, Berezhko, \& Baring 2000;
Berezhko, Ksenofontov, \& V\"{o}lk 2002;
V\"{o}lk, Berezhko, \& Ksenofontov 2003;
Aharonian \& Atoyan 1999; Reynolds 1998; Dyer et al. 2001;
Allen, Petre, \& Gotthelf 2001;
Achterberg, Blandford, \& Reynold 1998).
For example, at present the origin of TeV $\gamma$-rays 
can be explained by both leptonic and hadronic models.
This comes from an insufficient theoretical understanding;
apart from a globally successful picture of the DSA,
detailed but important processes,
such as the injection or the reflection of accelerated particles
that determines the above unknown quantities,
are not well understood.
Worse yet,
previous observations in the hard X-ray band had
insufficient spatial resolutions
to resolve small-scale structures near the shock front,
and could not strongly constrain the theoretical parameters.

Recently, Bamba et al. (2003a,b)
reported on the results 
for not only spectral but spatial studies of
 thermal and non-thermal shock structure in the NE rim of SN~1006
with the excellent spatial resolution of {\it Chandra}.
Similar results are also reported by Long et al. (2003)
with {\it Chandra} data.
Bamba et al. (2003a) estimated the scale length of thermal and 
non-thermal X-rays
in both upstream and  downstream of the shock front,
which means that the direct measurement of the diffusion
coefficients has become possible.

In this paper, 
we show that the important physical parameters for
the magnetic field in the acceleration site can be constrained 
by the spatial distribution of observed nonthermal X-rays.
\S~\ref{section:obs} summarized 
results of data analyses by Bamba et al. (2003a).
We construct two models in \S~\ref{sec:model};
one is assumed that the maximum energy of 
accelerated electrons are determined by the age of SNR,
while the other is by the energy loss process such as 
synchrotron or IC cooling.
Finally, \S~\ref{section:dis} is devoted to the  discussion 
on validity of our estimation.
Throughout the paper, indices
``u'' and ``d'' represent upstream and downstream, respectively. 

\section{Observed Properties of Nonthermal NE shell}
\label{section:obs}
In this section, we summarize the observed consequence of SN~1006,
which will be used in the following sections.

We used the {\it Chandra} archival data of the ACIS,
which has the spatial resolution of 0.5 arcsec on the aim-point,
on the NE shell of SN~1006 (Observation ID = 00732)
observed on July 10--11, 2000 with the targeted position at
(RA, DEC) = 
($15^{\rm h}03^{\rm m}51\fs6$, $-41^{\rm d}51^{\rm m}18\fs8$)
in J2000 coordinates.
Bamba et al. (2003) and Long et al. (2003) report that
there are very thin filaments in the hard X-ray band image,
which must be
the acceleration sites of high energy electrons.
Details of the observation and the analysis are found in these papers.

As shown in Fig.~2 of Bamba et al. (2003),
they made profiles of six filaments and found that
the upstream scale length $w_\u$ ranges between 0.01 and 0.1~pc,
while the downstream scale length $w_\d$ varies from 0.06 to 0.4~pc
using the exponential function
with adopted distance of 2.18~kpc (Winkler et al. 2003).
The mean values of $w_\u$ and $w_\d$ are
0.05~pc and 0.2~pc, respectively.

Bamba et al. (2003) also fitted the X-ray spectra for the 
six filaments with an {\it srcut} model in the XSPEC package
(Reynolds 1998; Reynolds \& Keohane 1999).
The radio spectral index of 0.57 was adopted from
the result of Allen et al. (2001).
As a result, the best-fit roll-off frequency 
$\NRO=2.6$ (1.9--3.3)$\times 10^{17}$~Hz was derived.
The quantity $\NRO$ is written in terms of magnetic field ($B$) 
and the maximum energy of accelerated electrons ($\EM$) as
(Reynolds \& Keohane 1999)
\begin{equation}
\nu_{\rm rolloff}=5\times10^{17}\ {\rm Hz} \
\left(\f{B}{10\MG}\right)
\left(\f{\EM}{100\TeV}\right)^2 \ .
\L{RollOff}
\end{equation}
Since most of the nonthermal X-ray photons are observed in downstream,
the synchrotron radiation is mainly due to the downstream region.
Therefore, it is possible to adopt $B$ in Eq.~(\ref{RollOff})
 with the downstream magnetic field $B_\d$.

\section{Interpretation of the Observed Width of 
Nonthermal X-ray Filaments}
\label{sec:model}
In this section, we
explain the observed scale length  of nonthermal X-ray
filaments.
Two simple models are considered in a context of
 DSA with a test-particle approximation.
We assume that the spatial distribution of 
nonthermal X-rays coincides with that of the accelerated electrons
to the  maximum energy, while thermal X-rays trace the
spatial profile of a background plasma and hence 
a magnetic field.
For a steady state, there is no spatial structure of accelerated 
particles in  downstream (Blandford \& Ostriker 1978).
However, one should consider the finite-time
or energy-loss effect, which makes the spatial profile 
in downstream.

For simplicity, we assume magnetic fields are 
spatially uniform both in the upstream and downstream at least in 
the nonthermal X-ray emitting region.
Since the fraction of magnetic pressure to the ram pressure
is estimated as
$(B^2/8\pi)/(m_{\rm H}nu_\s^2)\sim2\times10^{-5}(B/10\MG)^2$,
where we assume the number density of thermal plasma 
$n\sim$1~cm$^{-3}$ and the shock velocity 
$u_\s\sim3\times10^3$~km~s$^{-1}$,
the magnetic pressure does not affect the dynamics of SNR.
We therefore can adopt the self-similar solution derived by
Ratkiewicz, Axford, \& Mckenzie (1994).
Our assumption of spatially uniform magnetic
field is a good approximation
in the narrow range around the shock front.

Since the wide-band spectrum shows a break on the
X-ray band, electrons accelerated near the
maximum energy $\EM$ contribute the nonthermal X-ray emission.
The quantity $\EM$ is determined by the
age of SNR or by the balance of the acceleration and the energy-loss
efficiencies.
To discuss this, we consider three time scales;
the acceleration time scale $t_\ac$,
the energy loss time scale $t_\l$,
and the age of SN~1006 $t_\ag$.
The energy loss of high energy electrons can be neglected when 
$t_\ac<t_\l$.
Then,  $\EM$ is determined by $t_\ac=t_\ag$.
On the other hand, 
$\EM$ is limited by the energy loss if $t_\ac=t_\l<t_\ag$.
We investigate these two cases separately 
in the following subsections.

\begin{figure*}
\centering
\includegraphics[width=17cm]{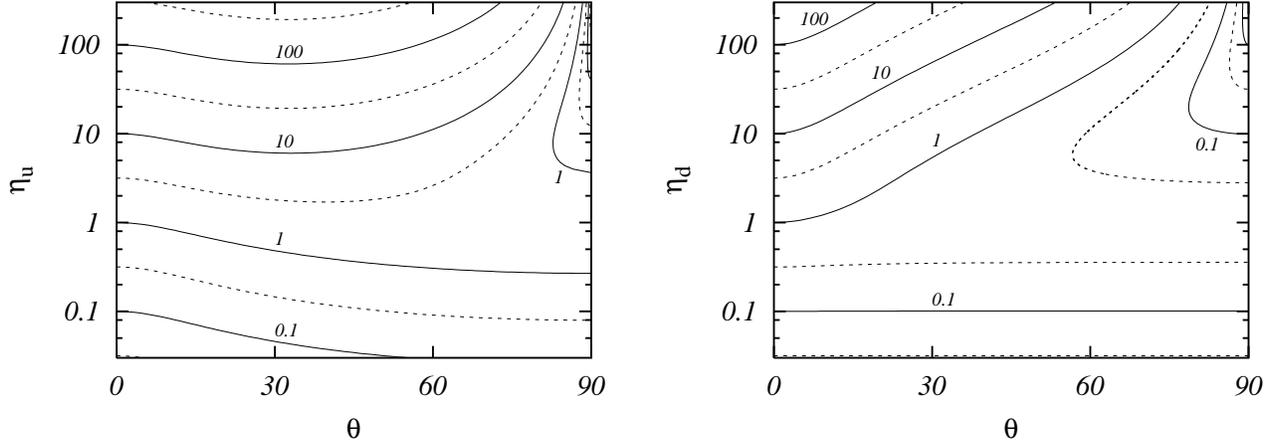}
\caption{
Contour maps of the functions $f(\eta_\u,\theta)$ (left panel)
and $g(\eta_\d,\theta)$ (right).
\label{fig:contour}}
\end{figure*}

The acceleration time is given by Drury (1983) as
\begin{equation}
t_\ac=\f{3}{u_\u-u_\d}\left(\f{K_\u}{u_\u}+\f{K_\d}{u_\d}\right) \ ,
\L{AccTime}
\end{equation}
where $u$ and $K$ are
the velocity of bulk flow in the shock frame 
and the diffusion coefficients for accelerated electrons
at the maximum energy, respectively.
The diffusion coefficients are given by quasi-linear theory
(Skilling 1975; Blandford \& Eichler 1987).
Let the mean free path of accelerated electrons parallel to the
magnetic field be a constant factor $\eta$ times the gyro radius
$r_{\rm g}=E/(\e B)$ in both upstream and downstream.
Then the diffusion coefficients in upstream and downstream can 
be written in terms of $\eta_\u$, $\eta_\d$, and the angle
between upstream magnetic field and the shock normal $\theta$
(Jokipii 1987),
\begin{equation}
K_\u=\f{c\EM}{3\e B_\u}\,\eta_\u\left(\cos^2\theta+\f{\sin^2\theta}
{1+\eta_\u^2}\right) \ ,
\L{DiffUp}
\end{equation}
\begin{eqnarray}
K_\d &=& \f{c\EM}{3\e B_\d} \,\eta_\d
(\cos^2\theta+r^2\sin^2\theta)^{-1} \N\\
&&\qquad\qquad\qquad 
\times\left(\cos^2\theta+\f{r^2\sin^2\theta}{1+\eta_\d^2}\right) \ , 
\L{DiffDown}
\end{eqnarray}
where $r$ is the compression ratio.
In this paper, since we assume that the shock is sufficiently strong
and that shock structure is not affected by the cosmic-ray pressure,
$r$ should be 4 and the upstream and downstream magnetic field are 
related as
\begin{equation}
\f{B_\d}{B_\u}=R(\theta)
:=(\cos^2\theta+r^2\sin^2\theta)^{\Half} \ .
\L{MagComp}
\end{equation}
Upstream and downstream velocities are related as
$u_\u=ru_\d\equiv u_\s$, and throughout the paper,
we adopt $u_\s=2.89\times10^3$~km~s$^{-1}$ (Ghavamian et al. 2002).
It is convenient to rewrite diffusion coefficients 
using Eqs.~(\ref{DiffUp}), (\ref{DiffDown}), and (\ref{MagComp})
as
\begin{equation}
K_\u=\f{c\EM}{3\e B_\d}f(\eta_\u,\theta) \ ,
\label{DiffUp2}
\end{equation}
\begin{equation}
K_\d=\f{c\EM}{3\e B_\d}g(\eta_\d,\theta) \ ,
\label{DiffDown2}
\end{equation}
where $f(\eta_\u,\theta)$ and $g(\eta_\u,\theta)$ are given by
\begin{equation}
f(\eta_\u,\theta)=
\eta_\u (\cos^2\theta+r^2\sin^2\theta)^{1/2}
\left(\cos^2\theta+\f{\sin^2\theta}
{1+\eta_\u^2}\right)\  ,
\label{DefF}
\end{equation}
\begin{equation}
g(\eta_\d,\theta)=
\eta_d (\cos^2\theta+r^2\sin^2\theta)^{-1}
\left(\cos^2\theta+\f{r^2\sin^2\theta}
{1+\eta_\d^2}\right) \ .
\label{DefG}
\end{equation}
Note that the downstream magnetic field $B_\d$ is used both in
Eqs.~(\ref{DiffUp2}) and (\ref{DiffDown2}).
Figure~\ref{fig:contour} shows the contour plots of
$f(\eta_\u,\theta)$ and $g(\eta_\d,\theta)$.
If $\theta\ga83\degr$, 
$f$ can be smaller than unity in the case of $\eta_\u>1$.
The quantity $g$ is smaller than 0.1 for $\eta_\d>1$
and $\theta\ga79\degr$.

Next, we consider the energy loss time scale.
Two processes may mainly cause the energy loss of electrons;
synchrotron radiation and IC effect by cosmic
microwave background photons.
In the following calculations, the latter process can be neglected.
Then, we can simply write
\begin{eqnarray}
t_\l &=& \f{6\pi m_\e^2c^3}{\sigma_TEB^2} \N\\
&=& 1.25\times10^{3}\ {\rm yrs} \
\left(\f{\EM}{100\TeV}\right)^{-1}
\left(\f{B}{10\MG}\right)^{-2} \ .
\L{SynTime}
\end{eqnarray}
When one compares $t_\l$ with $t_\ac$,
the mean value of the magnetic field
that accelerated electrons suffer should be adopted.
We can estimate the mean magnetic field as 
\begin{equation}
\langle B^2\rangle = \alpha B_\u^2 +(1-\alpha)B_\d^2 
= \chi B_\d^2\ ,
\end{equation}
where $\alpha=\Delta t_\u/(\Delta t_\u+\Delta t_\d)$ is the 
time fraction that accelerating electrons are in upstream.
For the particles in the acceleration process, we can estimate
$\Delta t_\u/\Delta t_\d\sim(K_\u/u_\u)/(K_\d/u_\d)=f/(rg)$
from Eqs.~(\ref{DiffUp2}) and (\ref{DiffDown2}).
Then, we obtain $\chi=(R^{-2}f+rg)/(f+rg)$,
which ranges between 0 and 1,
where $R$ is defined in Eq.~(\ref{MagComp}).

\subsection{Age-Limited Case}
\label{section:age}
First, we investigate the case in which
the acceleration time
is nearly equal to the age of SN~1006 $t_\ac\sim t_\ag$.
This condition implies that the observed nonthermal X-rays
are emitted by electrons that have been accelerated up to now.

\begin{figure*}
\centering
\includegraphics[width=17cm]{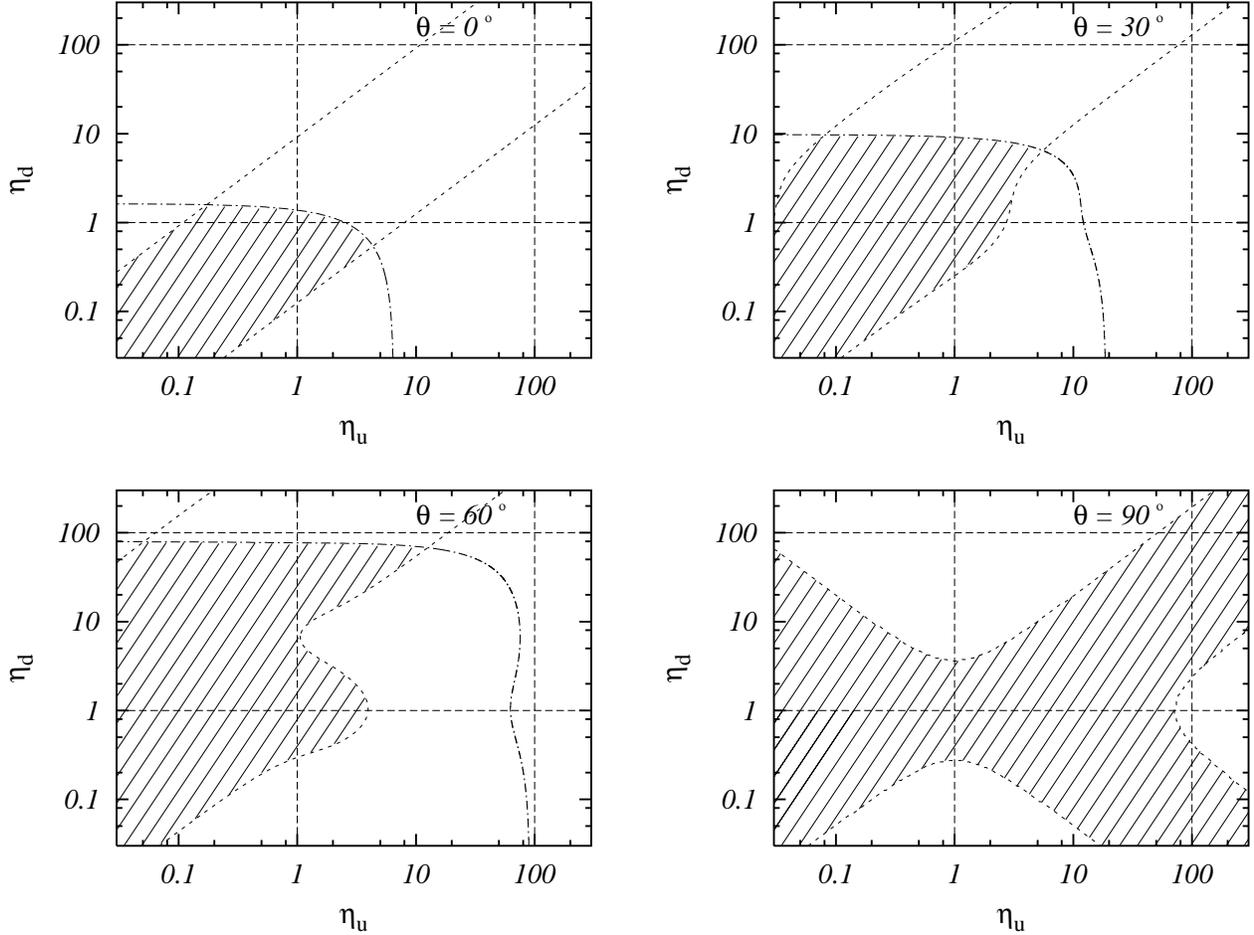}
\caption{
Points in the shaded regions surrounded by dotted and dot-dashed lines
satisfy Eqs.~(\ref{RollOff}), 
(\ref{MagComp}), (\ref{WuAge}), and (\ref{WdAge})
for fixed $\theta$ in the age-limited case.
Dotted lines represent Eq.~(\ref{fg1;age}) with
$w_d/w_u=0.6$ or 40, while
dot-dashed lines Eq.~(\ref{fg2;age}) with $\NRO=1.9\times10^{17}$~Hz,
the upper to which is forbidden since $t_\ac>t_\l$.
\label{fig:age}}
\end{figure*}

\begin{figure*}
\centering
\includegraphics[width=17cm]{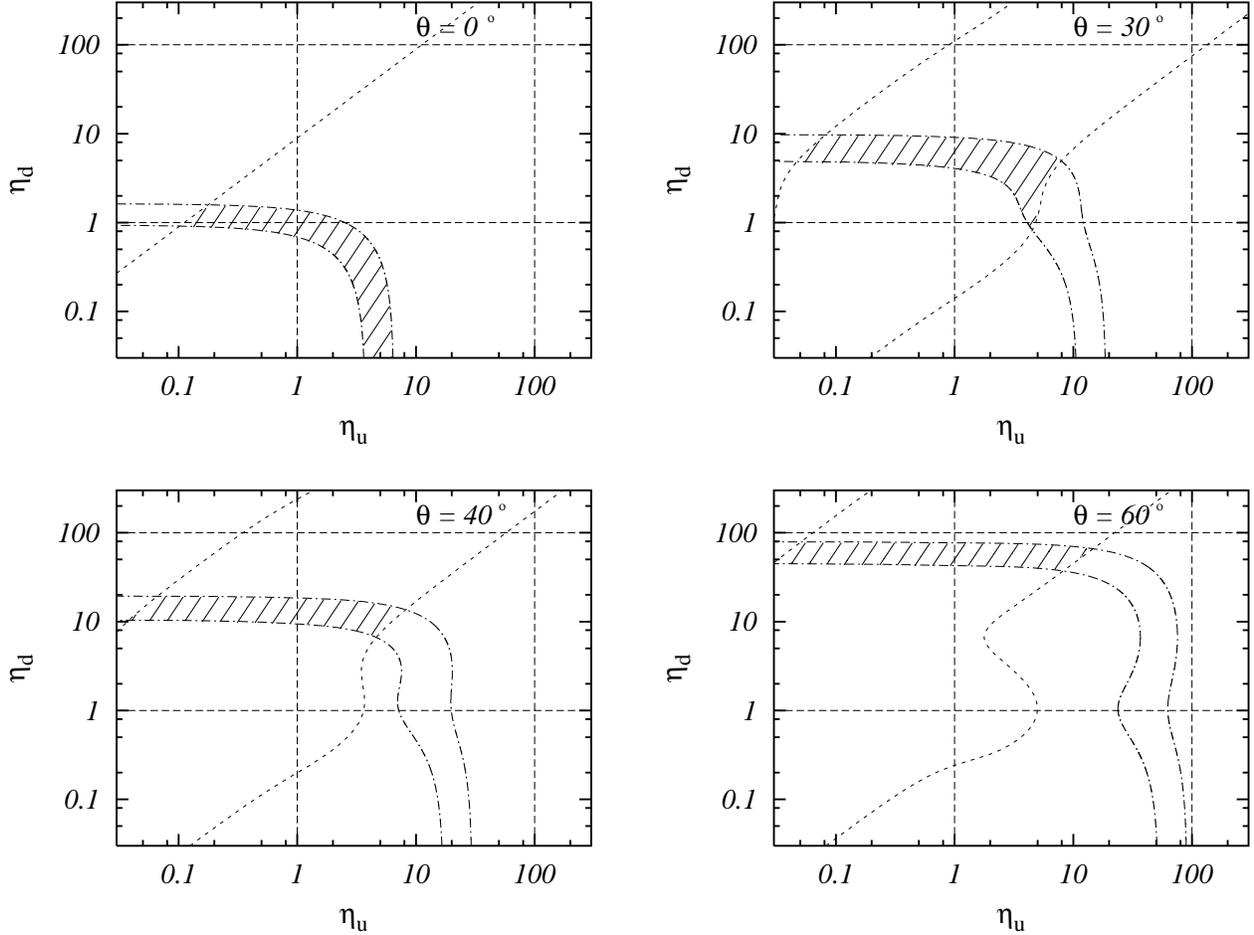}
\caption{
Points in the shaded regions surrounded by dotted and dot-dashed lines
satisfy Eqs.~(\ref{RollOff}), 
(\ref{MagComp}), (\ref{CoolLim}), (\ref{WuCool}), (\ref{WdCool}),
and $t_\l<t_\ag$
for fixed $\theta$ in the energy loss-limited case.
Dotted lines represent Eq.~(\ref{fg3;loss}) with
$w_d/w_u=0.6$ or 40, while
dot-dashed lines Eq.~(\ref{fg2;loss}) with 
$\NRO=1.9\times10^{17}$ or $3.3\times10^{17}$~Hz.
\label{fig:cool}}
\end{figure*}

\begin{figure*}
\centering
\includegraphics[width=17cm]{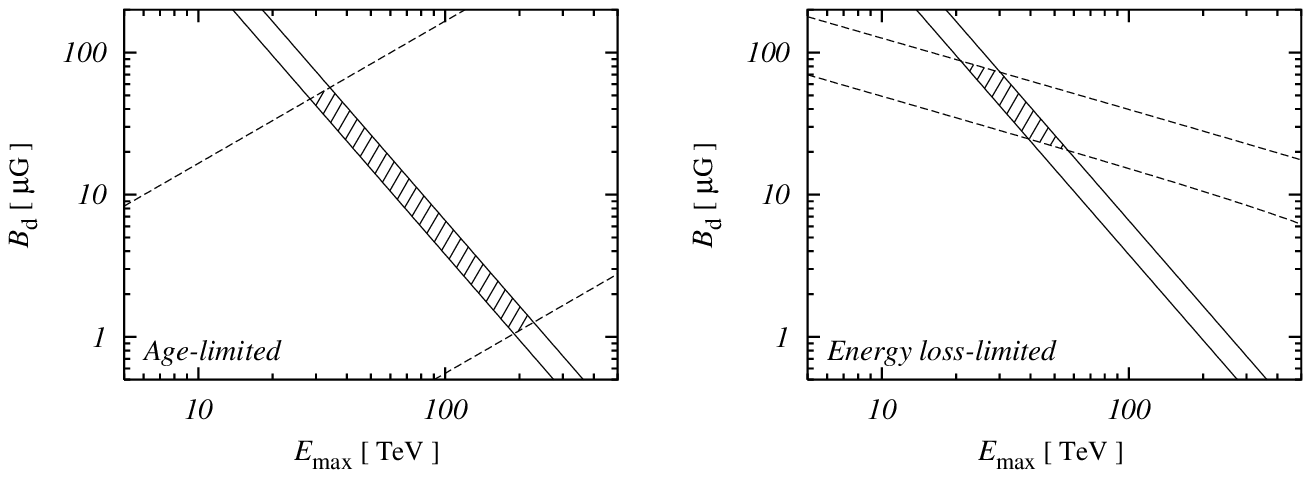}
\caption{The shaded areas indicate the most likely values of
magnetic fields just behind the shock front $B_d$ and
the maximum energy of shock-accelerated electrons $\EM$.
The left panel is for the age-limited case, while
the right the energy loss-limited model.
These areas are formed by the two solid lines for
a roll-off energy of $\NRO=(1.9$--3.3)$\times10^{17}$~Hz,
and the two dashed lines from
the observed width of nonthermal X-rays;
in the age-limited case,
a gyro-radius of accelerated electrons in the downstream 
should be $r_g=\EM/(\e B_\d)=(0.065$--20$)\times10^{-2}$~pc,
while in the energy loss-limited case,
cooling time of accelerated electrons
should be $t_\co=w_\d/u_\d=(0.81$--5.4$)\times10^2$~yrs.
\label{fig:EmaxBd}}
\end{figure*}

The particles in the diffusive shock acceleration process
are transported to the upstream by the diffusion,
and are advected to the downstream.
The diffusion and advection time scales to move a scale-length $w$
are $t_\ad=w/u$ and $t_\di=w^2/K$.
In the upstream, the accelerated particles can reach up to the 
place at which the advection is balanced with the diffusion,
i.e. $t_\ad=t_\di$.
Therefore, the observed upstream-width of nonthermal X-rays
can be written as
\begin{equation}
w_\u = \f{K_\u}{u_\u} 
=\f{c\EM}{3\e B_\d u_\s}f(\eta_u,\theta) \ .
\L{WuAge}
\end{equation}
In the downstream, if the particles are advected too far
from the shock front, they cannot return back to the upstream
to get further energy.
Therefore, the particles being accelerated should stay in the
region where $t_\di$ is smaller than $t_\ad$.
Thus, we obtain the same equation as  upstream,
\begin{equation}
w_\d = \f{K_\d}{u_\d}
=\f{c\EM}{3\e B_\d u_\s}rg(\eta_d,\theta) \ .
\L{WdAge}
\end{equation}
Then, Eq.~(\ref{AccTime}) becomes
\begin{equation}
t_\ac=\f{3r}{r-1}\f{w_\u+w_\d}{u_\s}
\sim 3.4\times10^2 \ {\rm yrs} \
\left(\f{w_\u+w_\d}{0.25\ {\rm pc}}\right) \ .
\label{AccTimeAge}
\end{equation}
Since $t_\ac$ is comparable to the age of the SN~1006, 
our assumption $t_\ac\sim t_\ag$ is justified.
Eliminating $\EM$ with Eq.~(\ref{RollOff}),
we can rewrite (\ref{WuAge}) and (\ref{WdAge}) as
\begin{eqnarray}
w_\u &=& 0.27\ {\rm pc}\
\left(\f{\NRO}{2.6\times10^{17}\ {\rm Hz}}\right)^{\frac{1}{2}}
\left(\f{B_\d}{10\ \MG}\right)^{-\frac{3}{2}}
f(\eta_\u,\theta)\ ,\N\\
&&
\label{WuAge2}
\end{eqnarray}
\begin{eqnarray}
w_\d &=& 1.1\ {\rm pc}\
\left(\f{\NRO}{2.6\times10^{17}\ {\rm Hz}}\right)^{\frac{1}{2}}
\left(\f{B_\d}{10\ \MG}\right)^{-\frac{3}{2}}
g(\eta_\d,\theta) \ ,\N\\
&&
\label{WdAge2}
\end{eqnarray}
respectively.

Let us take $w_\u\sim0.05$~pc, $w_\d\sim0.2$~pc,
and $\NRO\sim2.6\times10^{17}$~Hz as typical observed quantities
(Bamba et al. 2003a).
Then, Eqs.~(\ref{WuAge2}) and (\ref{WdAge2}) give
$B_\d\sim30f^{2/3}\ \mu{\rm G}\sim30g^{2/3}\ \mu{\rm G}$.
In the case of $\theta\sim0\degr$,
the value of $B_\d$ can be determined as follows.
We see that since $\eta_\u\geq1$,
$f$ should be greater than $\sim1$, {\it i.e.} $B_\d\ga30$~$\mu$G.
Independently of this,
the condition $t_\ac\la t_\l$ gives the upper limit of $B_\d$
as $B_\d\la30$~$\mu$G (see Eq.~[\ref{UpBdAge}]).
Therefore, we obtain $B_\d\sim30$~$\mu$G.
This argument has been done in Bamba et al. (2003a).
On the other hand, as we have mentioned before,
cases of $f\la1$ can be realized if $\theta\ga80\degr$,
{\it i.e.} the magnetic field is
nearly perpendicular to the shock normal.
Then $B_\d$ may be smaller than $\sim30$~$\mu$G.

Let us vary observed quantities $w_\u$, $w_\d$, and $\NRO$
in the range of the observed errors (at 90\% confidence level).
We have six unknown parameters $\EM$, $B_\u$, $B_\d$,
$\eta_\u$, $\eta_\d$, and $\theta$.
Conditions 
(\ref{RollOff}), (\ref{MagComp}), 
(\ref{WuAge}), (\ref{WdAge}), and $t_\ac\la t_\l$
are used to relate these quantities.
Details are summarized in \S~\ref{appendix}.
Figure \ref{fig:age} shows allowed regions of $\eta_\u$ or $\eta_\d$
for fixed $\theta$.
When we take $\theta=0\degr$, the case of
$\eta_\u=\eta_d=1$ (Bohm limit in both upstream and downstream) is
marginally acceptable, 
since $\eta$ should satisfy $1\le\eta\la c/u_\s\sim10^2$
(Jokipii 1987).
Then the magnetic field has values of $B_\u=B_\d=20$--78~$\MG$.
The equation~(\ref{UpBdAge}) shows that the
maximum value of magnetic fields is achieved when observed 
quantities $\NRO$, $w_\u$, and $w_\d$ have the minimum values.
On the other hand, when $\theta\ga85\degr$,
small magnetic fields are possible.
If we choose $w_\u=0.1$~pc, $w_\d=0.3$~pc,
and $\NRO=2\times10^{17}$~Hz, then
$\eta_\u\sim10$ and $\eta_\d\sim1$, 
the magnetic fields are $B_\d\sim4B_\u\sim14$--20~$\MG$.
This result has been suggested by Bamba et al. (2003a).

The left panel of Fig.~\ref{fig:EmaxBd}
 represents the allowed region of $\EM$ and $B_\d$
(see Eqs.~[\ref{EmaxAge}] and [\ref{BdAge}]).
The solid lines describe Eq.~(\ref{RollOff}) with
$\NRO=(1.9$--3.3$)\times10^{17}$~Hz, while the 
dashed lines a boundary of the region
in which a gyro-radius of accelerated electrons in the downstream
$r_{\rm g}=\EM/(\e B_\d)\propto g^{-1}w_\d$.
In order to satisfy $1\leq\eta\leq10^2$ 
and all of the other conditions, the quantity 
$g^{-1}(w_\d/{\rm pc})$ should range between 0.09 and 27.

\subsection{Energy Loss-Limited Case}
\label{section:cool}
If the maximum energy
of accelerated electrons $\EM$ is determined by
\begin{equation}
t_\ac=t_\l \ ,
\L{CoolLim}
\end{equation}
the motion of
accelerated particles toward the upstream might be obstructed 
by the energy loss effect as well as the advection.
Let us consider in the upstream, the energy loss time scale
$t_\co$\footnote{
To avoid the confusion, here we use the notation ``$t_\co$''
which is compared with $t_\ad$ or $t_\di$,
in order to distinguish ``$t_\l$'' used when one compares 
with $t_\ac$ or $t_\ag$.
}
in addition to the advection time scale 
$t_\ad=w_\u/u_\u$ and the diffusion time scale $t_\di=w_\u^2/K_\u$.
If $t_\di=t_\ad<t_\co$, the observed width of nonthermal
X-rays in the upstream is given by $w_\u=K_\u/u_\u$
as well as the age-limited case,
while in the case of $t_\di=t_\co<t_\ad$,
$w_\u$ is given by $w_\u=(K_\u t_\co)^{\Half}$.
Therefore, we can write
\begin{equation}
w_\u=\min\{K_\u/u_\u,\ (K_\u t_\co)^{\Half}\} \ .
\L{WuCool}
\end{equation}

On the other hand, the observed scale length of nonthermal X-ray 
filaments
in the downstream $w_\d$ is determined by the cooling time scale as 
\begin{equation}
w_\d=\max\{u_\d t_\co,\ (K_\d t_\co)^{\Half}\} \ .
\L{WdCool}
\end{equation}

We can now use five equations (\ref{RollOff}), (\ref{MagComp}),
(\ref{CoolLim}), (\ref{WuCool}), and (\ref{WdCool})
for six unknown parameters $\EM$, $B_\u$, $B_\d$,
$\eta_\u$, $\eta_\d$, and $\theta$,
and solve these equations with fixed $\theta$
under the condition of $t_\l<t_\ag$.
We summarize a detailed calculation procedure
in \S~\ref{appendix}.

Figure~\ref{fig:cool} shows the results for individual $\theta$.
We consider, as well as the age-limited case,
the errors associated with the analysis
of Bamba et al. (2003a).
When $\theta=0\degr$, the case of $\eta_\u=\eta_\d=1$ is
again  acceptable.
Then, the magnetic field is in the range of $B_\u=B_\d=23$--85~$\MG$.
If $\theta\la30\degr$,
the downstream magnetic field can be in the Bohm limit 
$\eta_\d=1$, then
$\eta_\u\sim1$--8 and $B_\d\sim23$--85~$\MG$.
However, if $\theta$ is larger than $\sim35\degr$,
then $\eta_\d\ga10$ and  $\eta_\u\la10$.
This implies that the upstream magnetic field is more turbulent
than the downstream, which seems to be unrealistic.

Indeed, as shown in \S~{\ref{sec:app2}},
scale lengths are given by $w_\u=K_\u/u_\u$ and $w_\d=u_\d t_\co$.
Then, $\EM$ and $B_\d$ are given by 
Eqs.~(\ref{EmaxCool}) and (\ref{BdCool}), respectively.
The right panel of Fig.~\ref{fig:EmaxBd} represents the 
allowed region of $\EM$ and $B_\d$ when $w_\d=0.06$--0.4~pc and
$\NRO=(1.9$--3.3)$\times10^{17}$~Hz.
All the points in the region satisfy other conditions.
Vink \& Laming (2003) discussed similar arguments for Cas~A.

\section{Discussion}
\label{section:dis}
We have argued the limitations on the model of DSA
from the recently observed scale length of 
nonthermal X-ray emitting region of NE shell of SN~1006
based on the assumption that
the observed spatial structure of nonthermal X-ray filaments
reflects that of accelerated electrons to the maximum energy.
A test particle approximation has been adopted, where
the back reactions of accelerated particles are neglected,
although one might have to consider corrections due to
nonlinear effects in order to explain the wide-band spectrum
from  radio to  TeV $\gamma$-rays.
Two models have been discussed; age-limited and
energy loss-limited models.
%
%
Note that in each model, the value and configuration of a magnetic 
field in the acceleration site can be discussed using only
the spatial distribution of the synchrotron X-rays,
which are emitted by accelerated electrons with energy several tens
of TeV, and the roll-off frequency, which is derived by the
fitting of the wide band spectrum from the radio to X-rays.

When a magnetic field is nearly parallel to the shock normal,
$\eta_\u$ and $\eta_\d$ should be nearly unity,
which means that the magnetic field is highly turbulent 
(near the Bohm limit),
and the magnetic field is in the range of 20--85~$\MG$.
Relatively strong parallel magnetic fields
are considered by several authors
(Berezhko et al. 2002; Ellison et al. 2000). 
Derived value of the magnetic field seems to be higher than the 
usual interstellar value of a few micro Gauss.
However, the mechanism pointed by Lucek \& Bell (2000)
may be able to
amplify the magnetic field.
These situation can be realized both in the 
age-limited and energy loss-limited model.

When the magnetic field is nearly perpendicular 
to the shock normal, relatively small values of
$B_\d=4B_\u=14$--20~$\MG$ are allowed, 
which can exist only in the age-limited model.
This value is consistent with the interstellar magnetic field
and assumed value in Bamba et al. (2003a) or other
papers (Allen et al. 2001; Reynolds 1998)
but slightly larger than that derived by CANGAROO observation
(Tanimori et al. 1998, 2001),
which assumes TeV $\gamma$-rays are emitted by the IC process.
This discrepancy may be solved if the filaments~1--6 are not the
main sites of TeV $\gamma$-rays generated by the leptonic process.
The observed flux of TeV $\gamma$-rays which are up-scattered CMB 
photons by synchrotron emitting electrons is written as
\begin{equation}
{\cal F}_{\rm IC}=2.5\times10^{-2}
\left(\f{B_\d}{20\ \mu {\rm G}}\right)^{-2}
{\cal F}_{\rm synch} \ ,
\end{equation}
where ${\cal F}_{\rm synch}$ is the observed flux of synchrotron
X-rays and estimated as 
$1.8\times10^{-12}$~ergs~s$^{-1}$~cm$^{-2}$ 
in the 0.5--10.0~keV band by \cite{bamba2003}.
Then, the contribution of the IC $\gamma$-rays
generated in the filaments~1--6 is only $\sim0.5$\%
of the whole flux coming from SN~1006.
This shows that the IC $\gamma$-ray emitting sites
have smaller magnetic field,  hence radiate less
synchrotron X-rays than in the filaments~1--6.
Or the TeV $\gamma$-rays might be emitted by the hadronic process.
In fact the most plausible position of the $\gamma$-ray emission
determined by CANGAROO is located north (and outside) of 
the filament~1 (Tanimori et al. 1998, 2001).
However, the region where the significance is higher than
half the maximum value extends over $\sim 0.2^{\circ}$,
which is almost the same as the standard deviation of the
point-spread function of the CANGAROO telescope. 
We need stereoscopic observations of TeV gamma-ray emissions 
by imaging atmospheric Cherenkov telescopes 
(for example, CANGAROO-III, HEGRA, and H.E.S.S.)
to establish the position and extent of the emission region.

We briefly discuss how our results may change if the test particle
approximation is dropped.
Although we should consider the spatial structure modified, 
for example, in the upstream precursor region,
we simply argue the cases of $r>4$ that are thought to be caused
 by nonlinear effects.
Indeed, Berezhko, Ksenofontov, \& V\"{o}lk (2002) showed that the
present value of total compression ratio is  about 6.
In the age-limited case, we observe that
the inferred magnitude of the upstream magnetic field
becomes smaller than that of the $r=4$ case
as $r$ is made larger (up to 7).  However,
the overall shapes and total areas of the allowed 
regions are changed only slightly for any case of $\theta$.
Thus the nonlinear effect would not be essential.
On the other hand, in the energy loss-limited case, the allowed region
in the $\eta_\u$--$\eta_\d$ plane becomes narrow and only the cases of 
$\theta=0^\circ$--10$^\circ$ can be exist for $4<r\la6$ while
no allowed region for $r\sim7$ case.
This comes from the fact that another restriction,
Eq.~(\ref{eq_coolFG}), emerges.
The larger $r$, the stronger the constraint  
because of the less efficient advection for fixed
shock velocity $u_\s$.
Furthermore, it can be shown that there exists no case in which
$w_\d$ is given by $(K_\d t_\co)^\Half$, since
the condition $t_\ac=t_\co$ is incompatible with 
the condition $u_\d t_\co<(K_\d t_\co)^\Half$ in the parameter
range of our interest.

It is important to determine the magnetic field configurations
in order to discuss the acceleration and/or injection efficiency.
In this paper, using the spatial distribution of nonthermal X-rays,
we have shown  that roughly two cases can exist;
high and parallel, and low and perpendicular magnetic field.
In the latter case, the back reaction of accelerated particles is 
small and thus a test-particle treatment is a good approximation.
The magnetic field amplification process  pointed by
Lucek \& Bell (2000) does not work well.
While in the former case, nonlinear effects are so efficient that
the magnetic field can be large.
The difference between these cases probably comes from the fact that 
the (proton) injection rate depends strongly on the shock obliquity
and diminishes as $\theta$ increases
(V\"{o}lk, Berezhko, \& Ksenofontov 2003).
In addition to our result, radio polarization data with high spatial
resolution may become further informations about the magnetic field
configuration.

In this paper, we have adopted the plane shock approximation.
For further details, it might be important to consider the 
curvature effect as discussed in 
Berezhko, Ksenofontov, \& V\"{o}lk (2003)
in order to produce more realistic scenario.

\acknowledgements 
Our particular thanks are due to T.~Nakamura, M.~Hoshino, and 
T.~Tanimori for their fruitful discussions and comments.
R.Y. and A.B. are supported by JSPS Research Fellowship 
for Young Scientists.
This work is supported by a Grant-in-Aid for Scientific Research,
No. 14340066 from the Ministry of Educations, Culture, Sports,
Science, and Technology of Japan, 
and also supported by a Grant-in-Aid for for the 21st Century COE
``Center for Diversity and Universality in Physics''.

\appendix
\section{Detailed calculations}
\label{appendix}
We summarize a procedure of practical calculations of $\eta_\u$
and $\eta_d$ from observed quantities.
Once $(f,g)$ is given for fixed $\theta$,
the allowed region of $(\eta_\u,\eta_\d)$ is derived by
Eqs.~(\ref{DefF}) and (\ref{DefG}).
Relations of $f$ and $g$  in age-limited case
and energy loss-limited case are different from each other.

\subsection{Age-limited case}
\label{sec:app1}
One can derive from Eqs.~(\ref{WuAge}) and (\ref{WdAge})
\begin{equation}
\f{g}{f}=r^{-1}\f{w_\d}{w_\u} \ .
\label{fg1;age}
\end{equation}
The condition $t_\ac\la t_\l$ gives the additional inequalities.
Substituting Eqs.~(\ref{DiffUp2}) and (\ref{DiffDown2}) into 
Eq.~(\ref{AccTime}), we derive
\begin{equation}
t_\ac=4\phi(f+rg)\f{c\EM}{3\e B_\d u_\s^2} \ ,
\end{equation}
where $\phi=\phi(r)=(3/4)r/(r-1)$.
Energy loss timescale that should be compared with $t_\ac$ 
is given by
\begin{equation}
t_\l=\f{6\pi m_\e^2c^3}{\sigma_T\EM B_\d^2}
\f{f+rg}{R^{-2}f+rg} \ .
\label{LossTime3}
\end{equation}
Using Eq.~(\ref{RollOff}), we can eliminate $\EM$ and $B_\d$,
and obtain
\begin{eqnarray}
R(\theta)^{-2}f+rg &\la& 4.8\
\phi^{-1}
\left(\f{\NRO}{2.6\times10^{17}\ {\rm Hz}}\right)^{-1} \N\\
&&\times
\left(\f{u_\s}{2.89\times10^8\ {\rm cm}\ \s^{-1}}\right)^2 \ .
\label{fg2;age}
\end{eqnarray}
The quantities $f$ and $g$ should satisfy Eqs.~(\ref{fg1;age})
and (\ref{fg2;age}).

The condition $t_\ac\la t_\l$ gives also
the upper limit of $B_\d$.
In the age limited case, one can derive
$\chi=(R^{-2}w_\u+w_\d)/(w_\u+w_\d)$.
Then using Eqs.~(\ref{SynTime}) and (\ref{AccTimeAge}), we derive
\begin{eqnarray}
B_\d &\la& 30\ \MG\ \phi^{-2/3}
\left(\f{\NRO}{2.6\times10^{17}\ {\rm Hz}}\right)^{-1/3} \N\\
&&\times
\left(\f{w_\u R^{-2}(\theta)+w_\d}
{0.25\ {\rm pc}}\right)^{-2/3}
\left(\f{u_\s}{2.89\times10^8\ {\rm cm}\ \s^{-1}}\right)^{2/3} \ ,
\N\\
&&
\label{UpBdAge}
\end{eqnarray}
where we eliminate $\EM$ with Eq.~(\ref{RollOff}).

Additionally, using Eqs.~(\ref{RollOff}) and (\ref{WdAge}), 
one can show $\EM$ and $B_\d$ can be expressed in terms of $g$ as 
\begin{eqnarray}
\EM &=& 41\ {\rm TeV}
\left(\f{r}{4}\right)^{-1/3}
\left(\f{\NRO}{2.6\times10^{17}\ {\rm Hz}}\right)^{1/3} \N\\
&&\times\left(\f{g^{-1}w_\d}{0.2\ {\rm pc}}\right)^{1/3}
\left(\f{u_\s}{2.89\times10^8\ {\rm cm}\ \s^{-1}}\right)^{1/3} \ ,
\label{EmaxAge}
\end{eqnarray}
\begin{eqnarray}
B_\d &=& 31\ \MG 
\left(\f{r}{4}\right)^{2/3}
\left(\f{\NRO}{2.6\times10^{17}\ {\rm Hz}}\right)^{1/3} \N\\
&&\times
\left(\f{g^{-1}w_\d}{0.2\ {\rm pc}}\right)^{-2/3}
\left(\f{u_\s}{2.89\times10^8\ {\rm cm}\ \s^{-1}}\right)^{-2/3} \ ,
\label{BdAge}
\end{eqnarray}
respectively.

\subsection{Energy loss-limited case}
\label{sec:app2}
For our parameters we consider, $w_\u$ and $w_\d$ are given by
\begin{equation}
w_\u=K_\u/u_\s \ ,
\label{WuLoss2}
\end{equation}
\begin{equation}
w_\d=u_\d t_\co \ .
\label{WdLoss2}
\end{equation}
We can calculate $\EM$ and $B_\d$ from Eqs.~(\ref{RollOff}) and
(\ref{WdLoss2}) as
\begin{eqnarray}
\EM &=& 39\ {\rm TeV} 
\left(\f{r}{4}\right)^{1/3}
\left(\f{\NRO}{2.6\times10^{17}\ {\rm Hz}}\right)^{2/3} \N\\
&& \times 
\left(\f{w_\d}{0.2\ {\rm pc}}\right)^{1/3}
\left(\f{u_\s}{2.89\times10^8\ {\rm cm}\ \s^{-1}}\right)^{-1/3} \ ,
\label{EmaxCool}
\end{eqnarray}
\begin{eqnarray}
B_\d &=& 35\ \MG 
\left(\f{r}{4}\right)^{-2/3}
\left(\f{\NRO}{2.6\times10^{17}\ {\rm Hz}}\right)^{-1/3} \N\\
&&\times
\left(\f{w_\d}{0.2\ {\rm pc}}\right)^{-2/3}
\left(\f{u_\s}{2.89\times10^8\ {\rm cm}\ \s^{-1}}\right)^{2/3} \ .
\label{BdCool}
\end{eqnarray}
Then, from Eqs.~(\ref{DiffUp2}) and (\ref{WuLoss2}), we derive
\begin{eqnarray}
f &=& 4.8\
\f{w_\u}{w_\d}\left(\f{r}{4}\right)^{-1}
\left(\f{\NRO}{2.6\times10^{17}\ {\rm Hz}}\right)^{-1} \N\\
&&\times
\left(\f{u_\s}{2.89\times10^8\ {\rm cm}\ \s^{-1}}\right)^{2} \ .
\label{fg1;loss}
\end{eqnarray}
The condition $t_\ac=t_\l$ gives
\begin{eqnarray}
R(\theta)^{-2}f+rg &=& 4.8\ \phi^{-1}
\left(\f{\NRO}{2.6\times10^{17}\ {\rm Hz}}\right)^{-1} \N\\
&& \times
\left(\f{u_\s}{2.89\times10^8\ {\rm cm}\ \s^{-1}}\right)^2 \ .
\label{fg2;loss}
\end{eqnarray}
Eliminating $\NRO$ from Eqs~(\ref{fg1;loss}) and (\ref{fg2;loss}),
we derive
\begin{equation}
\f{g}{f}=
\f{1}{4}\left[\phi^{-1}\f{w_\d}{w_\u}-
\left(\f{r}{4}\right)^{-1}R(\theta)^{-2}
\right] \ .
\label{fg3;loss}
\end{equation}
Equations~(\ref{fg2;loss}) and (\ref{fg3;loss}) determine $(f,g)$.

We can show that
the condition $t_\l<t_\ag$ is always satisfied.
Substituting Eqs.~(\ref{EmaxCool}) and (\ref{BdCool}) into
Eq.~(\ref{LossTime3}), we obtain
\begin{eqnarray}
t_\l &=& 2.7\times10^2\ {\rm yrs}\
\left(\f{r}{4}\right)
\left(\f{w_\d}{0.2\ {\rm pc}}\right) \N\\
&&\times
\left(\f{u_\s}{2.89\times10^8\ {\rm cm}\ \s^{-1}}\right)^{-1}
\f{f+rg}{R^{-2}f+rg} \ .
\end{eqnarray}
Then, $t_\l<t_\ag$ reduces
\begin{eqnarray}
\f{1+rg/f}{R^{-2}+rg/f} &<& 3.7
\left(\f{r}{4}\right)^{-1}
\left(\f{w_\d}{0.2\ {\rm pc}}\right)^{-1} \N\\
&&\times
\left(\f{u_\s}{2.89\times10^8\ {\rm cm}\ \s^{-1}}\right)
\left(\f{t_\ag}{1000\ {\rm yrs}}\right) \ .
\label{Relation:fg}
\end{eqnarray}
Using Eq.~(\ref{fg3;loss}), we rewrite
 the left hand side of Eq.~(\ref{Relation:fg}) as
\begin{equation}
\f{1+rg/f}{R^{-2}+rg/f}=
1+(1-R(\theta)^{-2})\,\phi
\left(\f{r}{4}\right)^{-1} \f{w_\u}{w_\d} \ .
\label{Relation:fg2}
\end{equation}
Substituting Eq.~(\ref{Relation:fg2}) into Eq.~(\ref{Relation:fg}),
we derive
\begin{eqnarray}
w_\d &+& (1-R(\theta)^{-2})\,\phi
\left(\f{r}{4}\right)^{-1}\,w_\u \N\\
&<& 0.74\ {\rm pc}\
\left(\f{r}{4}\right)^{-1}
\left(\f{u_\s}{2.89\times10^8\ {\rm cm}\ \s^{-1}}\right) \N\\
&& \times
\left(\f{t_\ag}{1000\ {\rm yrs}}\right) \ .
\end{eqnarray}
This equation is always satisfied since
$w_\u$ and $w_\d$ range
0.01--0.1~pc and 0.06--0.4~pc, respectively.

Finally in order to validate Eqs.~(\ref{WuLoss2}) and (\ref{WdLoss2}),
we confirm the conditions
\begin{equation}
K_\u/u_\s<(K_\u t_\co)^{\Half}\ ,
\label{eq:CondUp}
\end{equation}
\begin{equation}
u_\d t_\co>(K_\d t_\co)^\Half \ ,
\label{eq:CondDown}
\end{equation}
are always satisfied for the parameters of our interest.
Using Eqs.~(\ref{RollOff}), (\ref{DiffUp2}) and (\ref{SynTime}),
Eq.~(\ref{eq:CondUp}) can be rewritten as 
\begin{equation}
f<19.0
\left(\f{u_\s}{2.89\times10^8\ {\rm cm}\ \s^{-1}}\right)^2
\left(\f{\NRO}{2.6\times10^{17}\ {\rm Hz}}\right)^{-1} \ .
\end{equation}
This equation, together with Eq.~(\ref{fg1;loss}), reduces
\begin{equation}
\f{w_\u}{w_\d}<r \ ,
\end{equation}
which  is always satisfied since
$w_\u=0.01$--0.1~pc and $w_\d=0.06$--0.4~pc.
On the other hand, from Eqs.~(\ref{RollOff}), (\ref{DiffDown2}),
(\ref{SynTime}), and (\ref{eq:CondDown}), we derive
\begin{equation}
g<1.2 \left(\f{r}{4}\right)^{-2}
\left(\f{u_\s}{2.89\times10^8\ {\rm cm}\ \s^{-1}}\right)^2
\left(\f{\NRO}{2.6\times10^{17}\ {\rm Hz}}\right)^{-1} \ .
\label{eq_coolG}
\end{equation}
Combining this equation and Eq.~(\ref{fg2;loss}), we eliminate
$\NRO$ and $u_\s$ as 
\begin{equation}
\left[1-\left(\f{r}{4}\right)^{-1}\phi\right]g<
\f{1}{4}\left(\f{r}{4}\right)^{-2}\phi\,R(\theta)^{-2}\,f \ .
\label{eq_coolFG}
\end{equation}
As long as $r\leq4$, this is always satisfied since
$1-(r/4)^{-1}\phi=(r-4)/(r-1)\leq0$.

\end{document}